\documentclass[twocolumn,showpacs,preprintnumbers,superscriptaddress]{revtex4}
\usepackage{graphicx}

\begin{document}
\title{Phase Diagram of a Model for Diluted Magnetic Semiconductors\\
Beyond Mean-Field Approximations}

\author{Gonzalo Alvarez}
\affiliation{National High Magnetic Field Lab and Department of Physics, 
Florida State University, Tallahassee, FL 32310}

\author{Matthias Mayr}
\affiliation{Max-Planck-Institut f\"ur Festk\"orperforschung, 70569 Stuttgart, Germany.}

\author{Elbio Dagotto}
\affiliation{National High Magnetic Field Lab and Department of Physics, Florida State University, Tallahassee, FL 32310}

\date{\today}

\begin{abstract}
A lattice
spin-fermion model for diluted magnetic semiconductors (DMS) is investigated
numerically, improving on previously used mean-field (MF) 
approximations. Curie temperatures are
obtained varying the Mn-spin $x$ and hole $n$ densities, and the
impurity-hole exchange $J$ in units of the hopping amplitude $t$. 
Optimal values are found in the subtle intermediate 
regime between itinerant and localized carriers. Our main result is the
behavior of the Curie temperature at large $J/t$, where
a ``clustered'' state is observed and ferromagnetism is suppressed.
Formal analogies between DMS and manganites are also discussed. 
\end{abstract}


\maketitle

Diluted magnetic semiconductors  based on III-V compounds
are attracting considerable attention due to their combination
of magnetic and semiconducting properties, that may lead to
spintronic applications \cite{ohno,dietl}. 
Ga$_{1-x}$Mn$_x$As is the most studied of these compounds with
a maximum Curie temperature $T_C$$\approx$$110$~K at low doping $x$,
and with a carrier concentration $p$=$(n/x)$$<$$1$ due to
the presence of $\rm As$ antisite defects \cite{ohno} or Mn intersticials \cite{yu}.
It is widely believed that this ferromagnetism is carrier-induced,
with holes introduced by doping mediating the interaction between
$S$=5/2 Mn-spins. This Zener mechanism operates 
in other materials as well \cite{review}.

In spite of the excitement around 110~K DMS, 
room temperature ferromagnetism should be achieved
for potential applications,
with logic and memory operations in a single device.
For this reason, a goal of the present effort is
to analyze the dependence of $T_C$ on the parameters
$x$, $p$, and $J/t$, helping in setting realistic expectations for
DMS potential technological applications. This goal can only be achieved
with good control over the many-body aspects of the problem,
and for this purpose lattice Monte Carlo (MC) techniques are crucial,
improving on previously employed MF approximations.
Our results lead to an optimistic view in this respect,
since $T_C$ is found to increase linearly with $x$ up to $x$$\sim$0.25.

Our effort builds upon  previous
important DMS theoretical studies
\cite{dietl,macdonald,bhatt,dassarma,mayr,millis,calderon}. However,
to analyze whether $T_C$ can be substantially
increased from current values, techniques as
generic as possible are necessary.
In particular, both the strong
interactions and disorder must be considered
accurately, with computational studies currently providing  the best
available tools. 
For these reasons, our work differs from previous
approaches in important qualitative aspects:
(1) Some groups use a continuum six-band description of DMS
\cite{macdonald}. (2) Other theories assume
carriers strongly bounded to impurity
sites \cite{bhatt}, and employ Hartree-Fock 
approximations. 
(3) Dynamical MF theory (DMFT) \cite{millis} 
may not capture the percolative character of DMS, with a
random impurity distribution and cluster picture \cite{dassarma,mayr}.
(4) Other approaches use MF uniform states \cite{dietl}, or introduce
a reduced basis in simulations \cite{calderon}.
While the
previous work is important in describing current
DMS materials, our goal is to establish the phase diagram of a DMS
model avoiding MF approximations.

For the above mentioned reasons, 
here a generic MC study of a lattice spin-fermion
model for DMS is reported. The Hamiltonian is
\begin{equation}
{\hat H}=-t\sum_{<ij>,\sigma}{\hat c}^\dagger_{i\sigma} {\hat c}_{j\sigma} +
J\sum_{I}\vec{S}_I\cdot\vec{\sigma}_I,
\label{eq:ham}
\end{equation}
\noindent where 
${\hat c}^\dagger_{i\sigma}$ creates a hole at site $i$
with spin $\sigma$, and the hole spin operator interacting
antiferromagnetically with the localized Mn-spin $\vec{S}_I$ is
$\vec{\sigma}_I={\hat
c}^\dagger_{I\alpha}\vec{\sigma}_{\alpha,\beta}{\hat c}_{I\beta}$.
The carrier can visit $any$ site of the lattice 
(assumed cubic \cite{comment4}).
The interaction term is restricted to randomly selected 
sites, $I$, with a $S$=5/2  Mn-moment. 
These spins are here considered classical with $|\vec{S}_I|$=1, as widely
assumed, allowing for a MC simulation similarly as 
in other contexts \cite{assumption}. Photoemission
and band calculations locate $J/t$ near 2 \cite{okabayashi,calderon,millis,sanvito}.
Rather than fixing parameters
to current DMS values, here $J/t$ is unrestricted, while
$x$ and $p$ mainly vary in the range [0,1]. 
Approximations include the neglect
of on-site 
Coulomb repulsion, valid at small $x$ and $p$ where double occupancy
is unlikely. In addition, 
Mn-oxides investigations \cite{review} show that an intermediate or large
$J/t$ plays a role analogous to a Hubbard $U/t$ at {\it any} 
$x$ \cite{review}\cite{comment5}. 
At low $x$, the probability of nearest-neighbors (NN) Mn-spins is also 
low (0.0625 at $x$=0.25), justifying the neglect of an
antiferromagnetic (AF) Mn-Mn coupling. The 
hole motion is described by a one-band tight-binding model, while 
a more realistic model should include many bands as well as spin-orbit
interaction \cite{macdonald}. Despite this simplification,
our study considers the
underlying lattice, absolutely necessary 
for a qualitative understanding of the 
DMS phase diagram.

\begin{figure}[h]
\includegraphics[width=7.6cm]{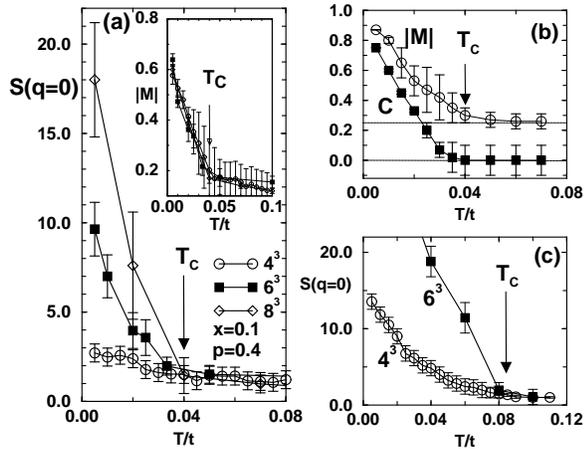}
\caption{(a) $S(q$=$0)$ vs. temperature, $T$, for $4^3$ (circles), 
$6^3$ (squares), and $8^3$ (diamonds) clusters at $J/t$=2.0, $x$=0.1 and
$p$=0.4, using the MC technique. The inset shows the magnetization
$|M|$ for the same clusters, 
with a vertical scale referred to the asymptotic $8^3$ $T$=$\infty$ value.
(b) $|M|$ (circles) and spin-spin correlation $C$
at maximum distance (squares)  
vs. $T$, on a $4^3$ cluster, at $J/t$=1.0, $x$=0.25, and $p$=0.4. 
The 0.25 horizontal line indicates the $|M|$  $T$=$\infty$ asymptotic 
value. (c) $S(q$=$0)$ vs. $T$,
at $J/t$=2.0, $x$=0.25, and $p$=0.4 using $4^3$ (circles) and 
$6^3$ (squares) clusters. In all cases $T_C$ is indicated. 
\label{fig:1}}
\end{figure}

The MC technique used here is as in Mn-oxides investigations \cite{review}:
it includes the full Exact Diagonalization (ED) of the hole sector for
each MC spin configuration, 
and density-of-states expansion calculations
\cite{furukawa}. The latter
allows us to reach clusters with up to $8^3$=512 sites if
up to 40 terms are included, reaching an accuracy comparable to ED for
smaller clusters. 
Both methods are nearly
exact, and the error bars of our results mainly arise from intrinsic
thermal fluctuations and averages over several random
Mn-disorder configurations. 
Comparing estimations of different clusters and based on
previous experience with similar models \cite{review}, $T_C$ can be calculated within a
$\sim$25\% accuracy, sufficient for our purposes \cite{calderon2}. 
The order parameter for the ferromagnetic-paramagnetic 
transition was taken to be
the absolute value of the magnetization of the Mn-spins
normalized to 1, namely 
$|M|$=$\frac{1}{xN}\sqrt{\sum_{I,R} \langle \vec{S}_I\cdot \vec{S}_R} 
\rangle$. 
Size effects are better visualized 
in the zero-momentum spin structure factor
$S(q$=$0)$=$\frac {1}{xN} \sum_{I,R} \langle \vec{S}_I\cdot\vec{S}_R 
\rangle$. Another useful quantity is
the spin-spin correlation at distance $d$, 
$C(d)$=$\frac{1}{N(d)}\sum_{|I-R|=d} \langle 
{{\vec{S}_{I}}\cdot{\vec{S}_{R}}} \rangle$, where $N(d)$ is the number
of pairs of Mn moments separated by a distance $d$.

Typical results for small and intermediate $J/t$
of our large-scale computational effort 
are in Fig.1a. There $S(q$=$0)$ and $|M|$
vs. temperature $T$ are shown
for three cluster sizes, $x$=0.1, and $p$=0.4.
Note the small size dependence of the
magnetization (inset), and the volume growth of $S(q$=$0)$ at fixed 
$T$$<$$T_C$. The estimated $T_C/t$ is $\sim$0.04, with
an uncertainty 0.01 sufficiently small for our purposes. 
Even with just the 4$^3$ cluster,
$T_C$ could be estimated fairly well, as shown in
Fig.1b. This is important to simplify our computational 
search for optimal $T_C$'s varying many parameters. 
In Fig.1b, the
temperature where a deviation from the high-$T$ limit is found is 
slightly larger than the $T_C/t$=0.04 obtained from larger clusters
(indicated). Studying the spin-spin correlation
at the largest available distance, a nonzero value characteristic of
an ordered ferromagnetic (FM) state was obtained at $T$ just below 0.04. 
Figure 1c provides another example of our
comprehensive $T_C$ study, using
just two cluster sizes at the $x$-$p$ location of our most optimal $T_C$, at
fixed $J/t$=2. Here the use of only 4$^3$ and 6$^3$ clusters provides once 
again a fairly accurate value $T_C$$\sim$0.08$t$.

To understand the qualitative 
$T_C$ trends, first consider the 
simplest case: the $p$ dependence at
fixed $J$ and $x$. Using the results in Fig.2a
contrasted against Fig.1b (same cluster size)
$T_C$ is found to change
by a factor $\sim$2, when $p$ varies from 0.1 to 0.4. However, this
tendency does not continue with increasing $p$, since at $p$=1 or beyond,
a FM state is not formed: the Pauli principle reduces
drastically the carrier kinetic energy, leading instead to an AF 
state. An example at $p$=3 and on an $8\times8$ cluster
(results are qualitatively similar in two and three dimensions)
is in Fig.2b, where the oscillations
in the spin correlations indicate staggered order. 
In general, the optimal $p$ is $\sim$0.5, between
the hole empty $p$=0 and saturated $p$=1 limits, as found with
DMFT \cite{millis}. A similar result
occurs in Mn-oxide models, recovered from
Eq.(1) at $x$=1. In that context, 
investigations at large Hund coupling, the analog of $J$ for DMS,
have shown that $p$=0.5 optimizes $T_C$ to a number $\sim$0.11-0.13$t$
\cite{review,alonso}, likely an 
{\it upper-bound} on the $T_C$ that could be achieved with Eq.(1).

\begin{figure}[h]
\includegraphics[width=7.6cm]{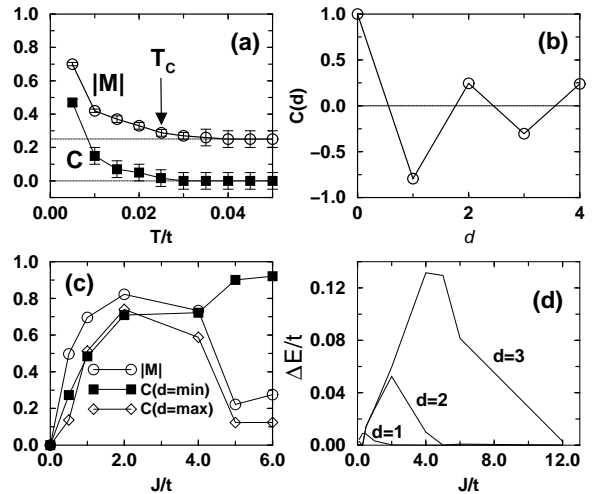}
\caption{(a) Magnetization $|M|$ (circles),
 and spin-spin correlation at maximum distance $C$ 
 (squares) vs. $T$, at $J/t$=1.0, $x$=0.25, and $p$=0.1, using a 
$6^3$ cluster. 
 (b) $C(d)$ vs. $d$ at $p$$\sim$3.0, $x$=0.25, $J/t$=1.0, and
$T/t$=0.005, using an $8^2$ cluster. The oscillations 
 indicate an AF state.
 (c) $|M|$ (circles), spin-spin correlation at minimum distance 
$C(d_{min})$ (squares), and at maximum distance $C(d_{max})$ 
(diamonds) vs. $J/t$,
 for a $4^3$ cluster at $x$=0.25, $p$=0.1, and $T/t$=0.005.
 (d) $\Delta E$= $E(\theta$=$\pi)$-$E(\theta$=$0)$ vs. $J/t$ 
calculated exactly on finite but large clusters
at $T$=0 for a system of 2 Mn-spins and 1 electron, 
$\theta$ being the relative angle between the Mn-spins. Results 
in 1, 2, and 3 dimensions are indicated. The spin distance 
is such that the associated effective $x$$\sim$0.1 is the same in all
cases. 
\label{fig:2}} 
\end{figure}

Consider now the $J/t$ dependence of $T_C$. The MF approximation
suggests $T_C^{mf}$$\propto$$J^2$. However, this does
not hold when more accurate methods are used in the calculations. In fact, 
for $J/t$$\rightarrow$$\infty$ and a Mn dilute system, the holes are trapped
in Mn-sites, 
reducing drastically the conductance and $T_C$. 
Small FM clusters of spins are formed 
at a temperature scale $T^*$, 
but there is no correlation between them, leading to a global vanishing
magnetization \cite{mayr}. These results $cannot$ be obtained within a
mean-field approximation. The large-$J/t$ ideas
can be tested in our MC simulation by monitoring
the short- and long-distance behavior of the spin-spin correlations $C(d)$. 
In a ``clustered'' state (large $J/t$), 
$C(d)$ at the shortest distance can be robust
at $T$$<$$T^*$, but $C(d)$ at the largest distance
vanishes due to the uncorrelated nature of the
magnetism between independent clusters (see Fig.2c). 
This subtle effect explains the
incorrect MF prediction, since $T_C^{mf}$$\sim$$T^*$, which
grows with $J/t$, rather than
the true $T_C$ (see also Fig.4a). 
Since both in the  $J/t$$\sim$0 and 
$J/t$=$\infty$ limits $T_C$ is suppressed, an
{\it optimal} $J/t|_{opt}$ must exist where $T_C$ is
maximized. Simulation results as in Fig.2c indicate that the
optimal $J/t$ value is close to 2.
This phenomenon is not captured in itinerant \cite{dietl} 
or localized \cite{bhatt} limits nor
by DMFT \cite{millis}, but it is observed in the present generic
MC simulations.

The existence of a $J/t|_{opt}$ can be illustrated just using
{\it two} spins and
{\it one} carrier in a finite cluster at $T$=0. For any 
fixed angle $\theta$ between the Mn-spins, assumed coplanar, the energy
is found exactly. The ground state of this $p$=0.5 system is always
at $\theta$=0 (FM-), while the energetically
worse state is $\theta$=$\pi$ (AF-configuration). Their energy
difference $\Delta E$
is a crude estimation of the FM state stability 
(Fig.2d). An optimal $J/t$ is found in
all dimensions, with stability increasing with the 
coordination number \cite{comment}. 
The result Fig.2d is understood measuring
the electronic density $n(i)$ of the same problem
on a chain (Fig.3a). At small $J/t$, the
 delocalization manifests in the nearly uniform density, 
leading to weak FM. At large $J/t$, strong localization
decouples the Mn-spins, producing again weak FM. 
However, there is an optimal
 value where the system takes advantage of $J/t$, but also allows for
a nonzero effective coupling among separated classical spins, leading 
to a stronger FM. 

\begin{figure}[h]
\includegraphics[width=7.6cm]{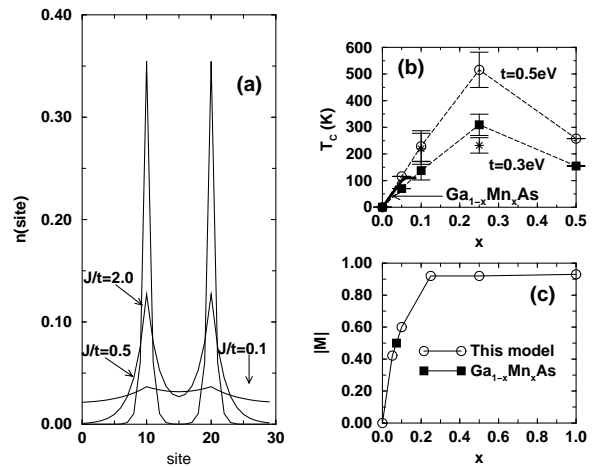}
\caption{(a) Exact $T$=0 local carrier
density, for 1 carrier and 2 parallel spins 
at sites 10 and 20, of a 30-site chain, varying $J/t$. 
(b) $T_C$ vs. $x$ for Ga$_{1-x}$Mn$_x$As \cite{potashnik} 
(thick line), and for model Eq.(1) using
 $t$=0.3 eV (squares) and $t$=0.5 eV (circles). In both cases,
$p$$\sim$0.4 and $J/t$=2.0. Typical error bars are shown. Stars are
results for $p$$\sim$0.5, $J/t$=1, and $t$=0.5 eV. 
(c) Magnetization $|M|$ vs. $x$ for model
Eq.(1) (circles) at $T/t$=0.005, $J/t$=2, $p$=0.4,
compared with the experimental value \cite{ohno} for
Ga$_{1-x}$Mn$_x$As at $x$$\sim$ 0.07 and $T$=2K (squares).
\label{fig:3}}
\end{figure}

Consider now the $x$ dependence of $T_C$. For
simplicity,  $J/t$$=$2 is mainly studied, which is both close to
optimal and experimentally realistic \cite{comment2}.
Fig.3b shows $T_C$ vs. $x$ at $p$$\sim$0.4, and for two reasonable values
of $t$. Experiments \cite{ohno} indicate a linear growth
of $T_C$ up to 5\% (shown), as in the numerical results. 
The slope of $T_C$ vs. $x$ is in remarkable agreement 
with MC predictions, in a reasonable range of $t$.
Regarding $x$$>$0.05, 
a reduction of $T_C$ was originally reported in experiments
\cite{ohno}. However,
recent data gathered with an optimized annealing treatment \cite{potashnik2}
indicate a $T_C$ ``plateau''. This still seems
in contradiction with the linearly growing $T_C$
of the MC results, but it suggests that even more
refined thin-films may continue increasing $T_C$ with increasing $x$. The MC
results clearly indicate linear behavior up to $x$$\sim$0.25 
(Fig.3b). To the extend that our model describes DMS quantitatively,
higher values of $T_C$ could be expected experimentally.
Regarding the presence of
a $T_C$ maximum at $x$=0.25: 
the origin of this effect is the growing probability with $x$ 
of having both holes and Mn-spins at NN-sites. 
In this case, AF links are formed since
$J/t$=2 is not so strong to keep the link FM,
reducing $T_C$ at large $x$
even at $p$=0.5. As $J/t$ grows, the effect diminishes and
the maximum in $T_C$ moves toward $x$=1, as 
naively expected. Reciprocally,
as $J/t$ decreases from 2 ($J/t$=1 shown in Fig.3b), 
the maximum in $T_C$ moves toward smaller $x$'s, and only $t$$>$0.5 can
provide high-$T$ ferromagnetism. This illustrates the key role that the
optimization of $J/t$ plays in these models, effect not captured
by MF approximations.

$|M|$ at $T$$\sim$0 is in Fig.3c. In
agreement with experiments, the 
$x$$\sim$0.1 result indicates a magnetization $\sim$50\% of its maximum value. 
This nonsaturated behavior originates in the random
distribution of Mn-spins, since Mn-clusters are formed providing a trap to
holes. Non-clustered spins are not much visited by those holes, and
their spins are not polarized. With growing $x$,
holes are more itinerant, polarizing the entire sample 
(Fig.3c) \cite{magnetic}.

\begin{figure}[h]
\includegraphics[width=6.3cm]{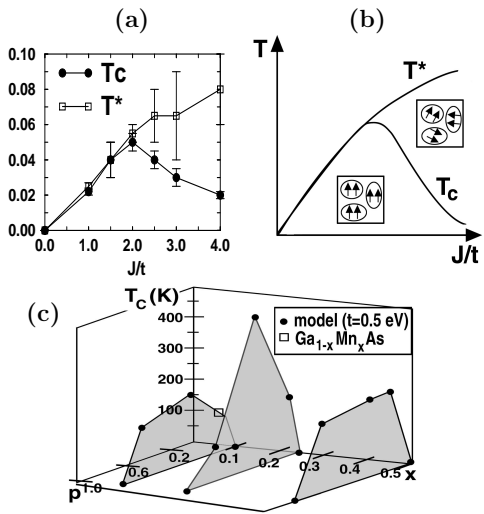}
\caption{(a) MC phase diagram in 2D  varying 
$J/t$, at fixed $x$ and $p$. At large $J/t$, 
a broad scale $T^*$ corresponds to the 
formation of uncorrelated clusters. $T_C$ is the ``true''
transition temperature, defined as the $T$ where 
FM correlations develop at the largest distance in the clusters used. 
At small $J/t$, those temperatures are similar. The
optimal $J/t$ is intermediate between itinerant and
localized regimes.
(b) Schematic phase diagram believed to be valid both in 2D and 3D,
with the clustered and FM states indicated.
(c) Numerically obtained $T_C$ vs. $x$ and $p$, 
at $J/t$=2.0. Filled
circles are from model Eq.(1) 
with $t$=0.5 eV. The open square corresponds to the
experimental value for Ga$_{1-x}$Mn$_x$As at $x$$\sim$0.1.
\label{fig:4}}
\end{figure}

In summary, MC investigations of a spin-fermion model
for DMS unveils substantial differences with previously
reported results employing MF techniques. The subtle regime of
intermediate $J/t$ appears the most relevant in these compounds.
$T_C$$\sim$0.08$t$ is an upper limit
for the FM critical temperature, result close to those
accepted for $x$=1 \cite{review,comment6}.
Our main results are summarized in Fig.4, that contain (a,b) the nontrivial 
$J/t$ dependence of $T_C$ and $T^*$, and (c) $T_C$ with varying $x$
and $p$, at optimal $J/t$. To the extend that the present model is 
applicable to DMS materials,
broad guidelines to improve $T_C$ can be established: 
(i) The optimal $J/t|_{opt}$$\sim$2 
must be intermediate between the itinerant and localized
limits (Fig.4a,b). This $J/t$, or larger, is expected to keep the
semiconducting nature of the state at $T$$>$$T_C$. Only band
calculations beyond our model can predict which particular material 
will have such an optimal $J/t$.
(ii) $x$ should be increased beyond 0.1. At $J/t|_{opt}$, the best
value is $x$$\sim$0.25.
Currently, $x$=0.14 is the experimental limit \cite{limit}.
(iii) The number of
antisite defects must be controlled 
such that $p$$\sim$0.5 ($p$$\sim$$1$ would be
detrimental due to competing antiferromagnetism).
(iv) As the coordination number grows, $T_C$
grows. (v) Finally, the simplest procedure to increase $T_C$ relies
on increasing the scale $t$. In fact, (Ga,Mn)As and
(In,Mn)As have different hybridization strengths \cite{oka}, and this
should be an important consideration in studying new materials.
Our work also suggests formal analogies between 
DMS and manganite models, with similar $T_C$'s, and a related clustered
state above ordering temperatures.

Work supported by NSF grant DMR-0122523 and by MARTECH.



\begin{thebibliography}{99}
\bibitem{comment4} 
Our study shows
the standard $T_C$ increase with coordination number. 
If a large $T_C$ is reached with cubic, even
larger values are expected with realistic FCC lattices.
\bibitem{assumption} This assumption is much used in Mn-oxide
studies, where a $S$=3/2 $t_{2g}$-spin
is usually considered classical \cite{review}.
\bibitem{comment5} Potential disorder is neglected \cite{dassarma,calderon}
(to be included in future publications).
\bibitem{calderon2} Our results for $x$=0.1, $p$=0.1, $J/t$=2, with a 
$T_C/t$$\sim$0.02 for an $8^3$ cluster, are comparable with results of
Ref.\cite{calderon} at $x$=0.1, $p$=0.07, $J/t$=3, using a $14^3$
cluster, that report $T_C/t$$\sim$0.03. This reasonable agreement in
the difficult regime of small $x$ and $p$ suggests that size effects are
under reasonable control in these studies.
\bibitem{comment} The optimal $J/t$ of this ``toy'' example is
larger only by a factor 2, compared with the realistic value Fig.2c.
\bibitem{comment2}
For $J/t$$\gg$1 values, the optimal
density for $T_C$ would be $x$=1, as in manganites.
\bibitem{magnetic}
Note also that the nonsaturated low-$T$ state can be fully polarized
with relatively small magnetic fields: in our studies, fields of just 10
T are sufficient to raise $|M|$ from 0.5 to nearly 1.0, at $x$=0.1.
\bibitem{comment6} 
$\rm La_{0.6} Sr_{0.4} Mn O_3$ has a $T_C$$\sim$340~K,
and our simulations suggest that similar values could be reached for DMS.
\end{thebibliography}
\end{document}